\documentclass[preprintnumbers,prd,showpacs,floatfix,preprintnumbers,amsmath,amssymb,nofootinbib,superscriptaddress]{revtex4}
\usepackage{graphicx}
\usepackage{epsfig}
\usepackage{bm}
\usepackage{amsfonts}
\usepackage{color}
\usepackage{subfigure}
\usepackage{amsmath}
\usepackage{amssymb}
\usepackage{graphicx}
\usepackage{makecell}
\usepackage{float}

\usepackage{amsmath,amssymb}

\begin{document}

\title{Bayesian Evidences for Dark Energy models in light of current observational data}

\author{Anto.~I.~Lonappan}
\email{antoidicherian@gmail.com}
\affiliation{Department of Physics, SB College\\ Changanassery, Kottayam 686101, Kerala, India.}
\affiliation{Department of physics, Presidency University, Kolkata, India}

\author{Sumit Kumar}
\email{sumit.k@icts.res.in}
\affiliation{International Centre for Theoretical Sciences, Tata Institute of Fundamental Research, Bangalore 560089, India}

\author{Ruchika}
\email{ruchika@ctp-jamia.res.in}
\affiliation{Centre for Theoretical Physics, Jamia Millia Islamia,
New Delhi-110025, India}

\author{Bikash R. Dinda}
\email{bikash@ctp-jamia.res.in}
\affiliation{Centre for Theoretical Physics, Jamia Millia Islamia,
New Delhi-110025, India}

\author{Anjan A Sen}
\email{aasen@jmi.ac.in}
\affiliation{Centre for Theoretical Physics, Jamia Millia Islamia,
New Delhi-110025, India}

\begin{abstract}
We do a comprehensive study of the Bayesian evidences for a large number of dark energy models using a combination of latest cosmological data from SNIa, CMB, BAO, Strong lensing time delay, Growth measurements, measurements of Hubble parameter at different redshifts and measurements of angular diameter distance by Megamaser Cosmology Project . We consider a variety of scalar field models with different potentials as well as different parametrisations for the dark energy equation of state. Among 21 models that we consider in our study, we do not find strong evidences in favour of any evolving dark energy model compared to $\Lambda$CDM. For the evolving dark energy models, we show that purely non-phantom models have much better evidences compared to those models that allow both phantom and non-phantom behaviours. Canonical scalar field with exponential and tachyon field with square potential have highest evidences among all the models considered in this work. We also show that a combination of low redshift measurements decisively favours an accelerating $\Lambda$CDM model compared to a non-accelerating power law model.
\end{abstract}

\maketitle

\date{\today}

\maketitle

\section{Introduction}

The concordance flat $\Lambda$CDM model is a simple yet successful description of our observable universe. Post Planck-2015, flat $\Lambda$CDM model  is consistence with data from Cosmic Microwave Background (CMB) \cite{ade}, from Supernova Type-Ia (SnIa) \cite{jla} as well as from the Baryon Acoustic Oscillations (BAO) measurements in galaxy surveys  \cite{bao} . But few recent low-redshift observations have shown tensions \cite{discrep,riess,kids,valen} with  $\Lambda$CDM model that is consistent with Planck-2015. Whether this is due to unknown systematics in both low and high redshift observations or due to physics beyond $\Lambda$CDM is still not confirmed. Moreover on theoretical side, $\Lambda$CDM model has its own problems like fine tuning and cosmic coincidence \cite{de}. This motivates people to consider dark energy beyond $\Lambda$ and to assume time evolution for dark energy. Recently, Zhao et al \cite{Zhao}, using combination of latest cosmological observations, have shown that dynamical dark energy model is preferred at $3.5 \sigma$ level.

The time evolution in dark energy is either modelled through different parametrisation for the dark energy equation of state ($ w = \frac{p}{\rho}$) or using different versions of evolving scalar field rolling over its potential. These are canonical scalar fields  \cite{quint}, non-canonical scalar fields \cite{noncan}, non-minimally coupled fields \cite{nonmin}, phantom fields \cite{phantom}, galileon fields \citep{gal,wali}  and many more.  These fields can also be classified as thawer or tracker/freezer \cite{lindcald} depending on the initial conditions as well as their late time evolutions.

Given any model, one can always constrain its parameters using available set of cosmological data. This exercise does not tell us which is a better model given a set of observational data. Calculating {\it Bayesian Evidence} \citep{gregory} is one of the ways to compare the probabilities of different models for given set of data. At present, when we have a reasonable set of very accurate cosmological measurements as well as a large set of dark energy models, {\it model comparison} using {\it Bayesian Evidence} can be very useful to pick reasonably better models with greater evidences from a large set of models describing the late time acceleration in the universe.

{\it Bayesian Evidence} in the context of cosmology was initially studied by Trotta \cite{trotta}. Later on, Martin et al. \citep{martin} have performed a comprehensive evidence calculation for a large set of inflationary models. In recent times, there have been few studies on {\it Bayesian Evidence} for dark energy models \citep{darkbayes}. In this work, we do a detailed study of {\it Bayesian Evidence} for a large class of cosmological models taking into account around $21$ different dark energy models. In this study, we consider the combination of CMB, SnIa, BAO, LSS, strong lensing, Megamsers and H(z) data. For LSS, we use the growth ($f\sigma_{8}$) measurements and the H(z) data includes the recent measurement of $H_{0}$ by Riess et al. ( hereafter R16) \citep{riess}. We also study how combinations of various set of data from cosmological observations influences the evidences of different models.

In section II, we describe different dark energy models that we consider in this work. In section III , we describe the observational data that we consider in the work and also the prior on different model and cosmological parameters. Section IV, we discuss the Bayesian evidences of different models and discuss the results of our study. In section V, we discuss the issue of accelerating vs non-accelerating Universe. Finally, we conclude in section VI.

\section{Theoretical Models}

In this section, we describe different dark energy models that we subsequently use for our evidence calculations. We consider a spatially flat FRW universe with scale factor $a(t)$.

\subsection{Canonical scalar field models}

We consider a minimally coupled scalar field with a Lagrangian \cite{quint}

\begin{equation}
 \mathcal{L} = \frac{1}{2}\partial_\mu\phi\partial^\mu\phi - V(\phi),
\end{equation}

\noindent
where $V(\phi)$ is the scalar field potential. The energy density and pressure for the scalar field are given by
\begin{eqnarray}
 \rho_\phi &=& \frac{1}{2}\dot{\phi}^2 + V(\phi) ,\\
 P_\phi &=& \frac{1}{2}\dot{\phi}^2 - V(\phi).
\end{eqnarray}
The equation of motion for scalar field  $\phi$ is
\begin{equation}
 \ddot{\phi} + 3H\dot{\phi} + V_\phi = 0,
\end{equation}

\noindent
where the subscript $\phi$ denotes the derivative with respect to field $\phi$. This type of scalar fields can be further divided in two subclasses: thawer and tracker/freezer. The freezing model has an initial fast roll phase when they track the background radiation or matter evolution; subsequently for specific choices of the potential $V(\phi)$, the equation of state decreases and asymptotically approaches $w=-1$ to initiate the late time acceleration. For the thawing models, the field is initially frozen at the flat part of the potential due to large Hubble friction and behaves like cosmological constant ($w=-1$). This initial condition can be achieved through inflation \citep{raghu}; subsequently the Hubble friction decreases and the field starts rolling and the equation of state slowly thaws away from the frozen $w=-1$ state towards higher values. 

In our evidence calculation, we use one particular tracking/freezing model with potential \citep{tiago}

\begin{equation}
V (\phi) = M^{4} [ e^{- \mu_{1} \frac{\phi}{M_{pl}}} + e^{- \mu_{2} \frac{\phi}{M_{pl}}} ]
\end{equation}

\noindent
where $ M $ is a constant having dimension of mass, $M_{pl}$ is Planck mass and $\mu_{1}, \mu_{2}$ are the two parameters in the model. We fix $\mu_{1} = 20$ to ensure the field initially track the matter. $\mu_{2}$ controls the equation of state at late times and it is a free parameter in our analysis. The parameter $M$ can be related to the energy density parameter of the field $\Omega_{\phi}$. In our calculation, we relate it to $\Omega_{\phi i}$ where the subscript ``i" denotes the value at decoupling ($z \sim 1100$) where we set our initial conditions. Hence $\mu_{2}$ and $\Omega_{\phi i}$ are the two model parameters for this tracking/freezing model.

For the thawer case, we consider four potentials: linear ($V(\phi) \sim \phi$), squared ($V(\phi) \sim \phi^2$), inverse-squared ($V(\phi) \sim \phi^{-2}$) and exponential ($V(\phi) \sim e^{\phi}$).  For thawer models, the parameters that determine the evolution of the field and its energy density for any potential are $\lambda = \frac{V_{\phi}}{V}$  and $\Omega_{\phi}$ (see  \citep{bobsen} for the relevant equations for thawing scalar fields). The two parameters in thawer model for any potential are $\lambda_{i}$ and $\Omega_{\phi i}$ where the subscript ``i" denotes the values at decoupling ($z \sim 1100$) as in the freezing case mentioned above.

\subsection{Non Canonical Scalar field models}

There are different non canonical scalar field models that have been considered as dark energy models. In our study, we use the non canonical models with Dirac-Born-Infeld (DBI) action ( This is also called Tachyon model in the literature) \citep{tach}:

\begin{equation}
  \mathcal{L} = - V(\phi) \sqrt{1 - \partial^\mu\phi\partial_\mu\phi}.
\end{equation}

\noindent
In natural units, the dimension of the tachyon field is $[Mass]^{-1}$. The energy density and pressure for the tachyon field are given by
\begin{eqnarray}
 \rho_t &=& \frac{V(\phi)}{\sqrt{1 - \dot{\phi}^2}},\\
 P_t &=& -V(\phi) \sqrt{1 - \dot{\phi}^2},
\end{eqnarray}

\noindent
The equation of motion for Tachyon field is
\begin{equation}
 \ddot{\phi} + 3H\dot{\phi}(1-\dot{\phi}^2) + \frac{V_{\phi}}{V}(1-\dot{\phi}^2) =0.
\end{equation}

For tachyon case, we assume thawing type initial conditions and consider the same four potentials as thawing scalar field.

\subsection{Galileon Models}

We consider the lowest non-trivial order of cubic Galileon Lagrangian along with a potential \citep{gal,wali}:

\begin{equation}
\mathcal{L} =  \frac{1}{2}(\nabla \phi)^2\Bigl(1+\frac{\alpha}{M^3}\Box \phi\Bigr) - V(\phi) ,
\label{eq:action}
\end{equation}

\noindent
where $\alpha$ is a  dimensionless constant; for $\alpha=0$, the Lagrangian~\eqref{eq:action} reduces to that of a canonical scalar field. $V(\phi)$ is the potential. $V(\phi) \sim \phi$ preserve the shift symmetry.  $M$ is a constant of mass dimension one; by a redefinition of the parameter $\alpha$, we can fix $M=M_{\rm{pl}}$. Action~\eqref{eq:action} can also be thought as a particular form of the Kinetic Gravity Braiding action \citep{braid}. The energy density and pressure of the galileon field are given by:

\begin{equation}
\rho_{g} = \frac{\dot{\phi}^2}{2}\Bigl(1-6\frac{\alpha}{M_{\rm{pl}}^{3}} H\dot{\phi}\Bigr) + V{(\phi)},
\end{equation}

\begin{equation}
p_{g} = \frac{\dot{\phi}^2}{2}\Bigl(1+2\frac{\alpha}{M_{\rm{pl}}^{3}}\ddot{\phi}\Bigr) - V(\phi),
\end{equation}
 
\noindent
where over dot is the derivative w.r.t. the time. Varying the action of the galileon field w.r.t the field $\phi$, we get the equation of motion for the field

\begin{equation}
\ddot{\phi} + 3H\dot{\phi}-3\frac{\alpha}{M_{\rm pl}^{3}} \dot{\phi}\Bigl(3H^2\dot{\phi}+\dot{H}\dot{\phi}+2H\ddot{\phi}\Bigr)+ V_{\phi}=0.
\end{equation}

\noindent
For galileon model also, we assume the thawing type of initial conditions at decoupling $z \sim 1100$. We introduce one new parameter $\epsilon = -6 \beta H^{2} \Big{(} \frac{d \phi}{d N} \Big{)}$, where $ \beta = \frac{\alpha}{M_{\rm pl}^{3}} $. The parameters of the galileon model are now, $\epsilon_{i}$, $\lambda_{i}$ and $\Omega_{\phi i}$  where the subscript ``i" denotes the values at decoupling  ($z \sim 1100$) and $\lambda =  \frac{V_{\phi}}{V}$ (see  \citep{wali} for the relevant equations for thawing galileon fields). Also in this case, we consider four types of potentials as in thawing scalar  and tachyon fields. Out of these four potentials, only linear potential respects the galileon symmetry and the rest of the potentials are purely phenomenological.

For all three dark energy models described above, the expression for Hubble parameter is given by

\begin{equation}
H^{2}(z) = H_{0}^2 \left(\Omega_{m0} (1+z)^{3} + \Omega_{r0} (1+z)^{4} + \Omega_{\phi 0} f(z)\right),
\end{equation} 

\noindent
where $f(z)$ is a function of dark energy energy density $\rho_{\phi}$ for canonical scalar field, $\rho_{t}$ for tachyon and $\rho_{g}$ for galileon. $H_{0}$ ($100h \hspace{1mm} Km/sec/Mpc$) is the present day Hubble parameter. Here $\Omega_{m0}$, $\Omega_{r0}$  and $\Omega_{\phi 0}$ are the present day density parameter for matter (that includes baryons and cold dark matter), radiation and scalar field respectively with the flatness condition $\Omega_{m0} + \Omega_{r0} + \Omega_{\phi 0} =1$.   So any two of the density parameters are independent. In our study, we take the density parameter for radiation and scalar field as independent ones and that for matter will be a derived parameter. Moreover, both $\Omega_{r0}$ and $\Omega_{\phi 0}$ can be related to their values at decoupling. We already mention above that $\Omega_{\phi i}$ is a model parameter for all three scalar field models. We also consider $\Omega_{r i}$ as another free parameter in our evidence calculations.

\subsection{Dark Energy Parametrisation}

Given a large number of scalar field models with a variety of potentials, it is always difficult to test all the individual models. Instead, we often use a parametrisation of dark energy evolution that broadly describes a large number of scalar field dark energy models. Parametrising the dark energy equation of state ($w = p/\rho$) is the most common practice  and there are also a large number of parametrisations for $w(z)$ available in the literature. Here we use the CPL \citep{cpl}, GCG \citep{gcg}, BA \citep{ba}, JBP \citep{jbp} and 7CPL \citep{7cpl} parametrisations for our evidence calculations together with standard $\Lambda$CDM and constant dark energy equation of state model: 

\begin{eqnarray}
w(z) &=& -1 \hspace{3mm} (\Lambda CDM)\nonumber\\
w(z) &=& w_{DE} (constant) \hspace{3mm} (wCDM)\nonumber\\
w(z) &=& w_{0} + w_{a}\frac{z}{1+z} \hspace{3mm} (CPL) \hspace{1mm}\nonumber\\
w(z) &=& -\frac{w_{0}}{w_{0} + (1-w_{0}) (1+z)^{3 (1+w_{a})}} \hspace{3mm} (GCG) \hspace{1mm}\\
w(z) &=& w_{0} + w_{a} \frac{z(1+z)}{1+z^2} \hspace{3mm} (BA) \hspace{1mm}\nonumber\\
w(z) &=& w_{0} +w_{a}\frac{z}{(1+z)^2} \hspace{3mm} (JBP) \hspace{1mm}\nonumber\\
w(z) &=& w_{0} + w_{a} \left(\frac{z}{1+z}\right)^7 \hspace{3mm} (7CPL) \hspace{1mm}\nonumber
\end{eqnarray}

\noindent
All the parametrisation, except the first two, contain two parameters $w_{0}$ and $w_{a}$ where $w_{0}$ is related to the present value of the equation of state for the dark energy and $w_{a}$ determines its evolution with time. Except GCG and 7CPL, the rest of the non-constant parametrisations represent the thawing evolution much better than the tracking evolution. In contrast, 7CPL fits the tracking/freezing model better \citep{7cpl}. The GCG mimics freezing evolution for $(1+w_{a}) > 0$ whereas for $(1+w_{a}) < 0$, it mimics the thawer evolution. We call them GCG-tracker and GCG-thawer respectively. With this, for a spatially flat FRW universe, the expression for Hubble parameter becomes

\begin{equation}
H^{2}(z) = H_{0}^2 \left(  \Omega_{m0} (1+z)^{3} + \Omega_{r0} (1+z)^{4} +  (1-\Omega_{m0} - \Omega_{r0}) \exp\left[3 \int \frac{1+w}{1+z} dz \right]   \right).
\end{equation}

Once we know the expression for the Hubble parameter $H(z)$, we can calculate all the observables related to the background cosmology.  The growth of matter fluctuations on sub-horizon scales and in linear  regime is given by 

\begin{equation}
\ddot{\delta_{m}} + 2 H \dot{\delta_{m}} - 4\pi G \bar{\rho_{m}} \delta_{m} = 0,
\end{equation}

\noindent
where $\delta_{m}$ is the matter density contrast, $\bar{\rho_{m}}$ is the background matter energy density and ``{\it over dot}'' represents the derivative with respect to the cosmological time $t$. $H$ is the Hubble parameter given by either (14) or (16). Here we use the Newtonian approximation which is valid on sub-horizon scales and also ignore the dark energy perturbations on sub-horizon scales. One can also define linear growth rate as:

\begin{equation}
f = \frac{d\log\delta_{m}}{d\log a}.
\end{equation}

The quantity $f(z)\sigma_{8}(z)$,  where $\sigma_{8}(z)$ is the rms fluctuation of linear over density field $\delta_{m}$ within a box size of $8 h^{-1}$ Mpc, is a model independent estimator of the observed growth history in the universe. On sub-horizon scales ignoring dark energy perturbations, one can write \cite{luca} 

\begin{equation}
f(z) \sigma_{8}(z) = \sigma_{8} \frac{\delta_{m}^{'}}{\delta_{m}(z=0)},
\end{equation}

\noindent
where ``{\it prime}'' denotes differentiation with respect of $\log (a)$ and $\sigma_{8} = \sigma_{8} (z=0)$. 

\section{Data and Prior}

To calculate the {\it Bayesian Evidence} for each models, we use the following observational data for our analysis:

\begin{itemize}

\item 

The measurements of the luminosity distance of SNIa from the ``{\it Joint Light Curve Analysis (JLA)}''  taken from SDSS and SNLS catalogue \cite{jla}.

\item 
We use the isotropic BAO measurements from 6dF survey \cite{6df} at $z=0.106$, SDSS data release for main galaxy sample (MGS) \cite{mgs} at $z=0.15$ and eBoss quasar clustering \cite{eboss} at $z=1.52$. We also use the anisotropic measurements from BAO only analysis of BOSS  analysis \cite{boss} and Lyman-$\alpha$ forest samples \cite{ly} at redshifts $z=0.38$, $z=0.61$ and $z=2.4$. For all these measurements and the corresponding covariance matrix, we refer readers to the recent work by Evslin et al \cite{jarah}.  

All the BAO measurements involve the BAO length scale $r_{d}$ which is the sound horizon at drag epoch. In our analysis, we keep $r_{d}$ as a free parameter.

\item
Angular diameter distances measured using water megamasers under the Megamaser
Cosmology Project \cite{maser}

As both Megamasers cosmology project and different BAO measurements measure the angular diameter distances or its different combinations at different redshifts, in our subsequent analysis, we refer different BAO measurements and the measurements from Megamasers Cosmology Project in a combined head "BAO" unless specifically mentioned.

\item We also use the collection of strong lensing time-delay measurements by H0LiCOW experiment \cite{holicow}. We denote this as "TDSL".

\item 

The measurement of $f(z)\sigma_{8}(z)$ by various galaxy surveys as compiled by Basikalos et al \cite{growth}. We denote this as "GROWTH".

\item

The acoustic scale and CMB shift parameters as measured by Planck-2015 observations \cite{ade1}. We denote this as "CMB".

\item

The measurement of Hubble parameter as a function of redshift as compiled by Farooq et al \cite{hubble}.

\item

The latest measurement of $H_{0}$ by Reiss et al \cite{riess} (Riess16).

In our subsequent analysis, we use "H(z)" to denote the $H$ data at different redshifts including the $H_{0}$ measurement unless specifically mentioned.

\end{itemize}

We use {\it flat prior} for both model and cosmological parameters in our analysis. In Table I, we list prior for different parameters used in our evidence calculations. 

\begin{widetext}
\begin{center}
\begin{table}[t]
\centering
\caption{Parameters used in different models and their prior. }
\begin{tabular}{ccc}
\Xhline{3\arrayrulewidth}
\textbf{Parameter} & \textbf{Models} & \textbf{Prior (uniform)}\\ 
\Xhline{3\arrayrulewidth}
$\Omega_{m0}$ & $\Lambda$CDM, $w$CDM, CPL, GCG,BA,JBP,7CPL & [0.1, 0.9]\\
h & All & [0.5, 0.9]\\
$\Omega_{r0}$ & $\Lambda$CDM, $w$CDM, CPL, GCG,BA,JBP,7CPL & [3.0, 9.0] $\times 10^{-5}$\\
$\sigma_8$ & All & [0.6, 1.0]\\
$r_{d}$ & ALL & [130,170]\\
$w_{DE}$ & $wCDM$,  & [-1.8, -0.4]\\
$w_0$ & CPL,BA,JBP,7CPL &[-1.9, -0.4]\\
$w_0$ & GCG-thawer, GCG-Tracker &[0.5, 1.0]\\
$w_a$ & CPL,BA,JBP,7CPL & [-4.0, 4.0]\\
$w_a$ & GCG-thawer & [-3.0, -1.01]\\
$w_a$ & GCG-tracker & [-0.99, 1.0]\\
$\Omega_{\phi i}$ & Scalar tracker, scalar thawer, Tachyon, Galileon   & [0.8, 3.0] $\times 10^{-9}$\\
$\mu_{2}$ & scalar tracker & [0,1]\\
$\epsilon_{i}$ & Galileon & [0.0, 50.0]\\
$\lambda_i$ & scalar thawer, Tachyon, Galileon &[0.2, 1.0]\\
$\Omega_{r i}$ & Scalar tracker, scalar thawer, Tachyon, Galileon  & [0.09, 0.17]\\
\Xhline{3\arrayrulewidth} 
\end{tabular} 
\end{table}
{\label{prior}}
\end{center}
\end{widetext} 

\section{Bayesian Evidence and Results}
In Bayesian data analysis, Bayes theorem \citep{gregory} relates the probability of parameters $\Theta_M=\{\theta^i_{M}; i=1,2,....W\}$ describing a model $M$ with the data $D$ and prior information $I$. According to Bayes' theorem:
\begin{equation}\label{Bayes}
\textit{P}(\Theta_M |D,I) = \frac{\textit{P}(D|\Theta_M, I)\times \textit{P}(\Theta_M | I)}{\textit{P}(D|I)}
\end{equation}
where $\textit{P}(\Theta_M |D,I)$ is the posterior probability of model parameter $\Theta_{M}$  given a data $D$. $\textit{P}(D|\Theta_M, I)$ is known as likelihood function which tells the probability of the realisation of data $D$ given a model M (along with model parameters). $\textit{P}(\Theta_M| I)$ is the prior probability function on the parameters of the model. The quantity $\textit{P}(D|I)$ is called the marginal likelihood or the evidence of the model M given the data D.

In parameter estimations of models, we usually ignore evidence for computing posterior as it does not affect the shape of posterior distribution. But in model selection, evidence plays the crucial role and is used to normalize the posterior over $\Theta_{M}$. So {\it Bayesian Evidence} of a model with \textit{W} dimensional parameter space is given by,
\begin{equation}\label{Z_int}
\mathcal{Z} = \int{\textit{P}(D|\Theta)\times \textit{P}(\Theta)}d^W\Theta.
\end{equation} 
\noindent
Models with higher {\it Bayesian Evidence} are favoured over those with lower evidence. We use Jeffreys' scale \citep{jeff} to compare models with different ${\mathcal{Z}}$ as shown in Table II:

\begin{widetext}
\begin{center}
\begin{table}[H]
\centering
\caption{table}{Jeffrey's Scale for Evidence}
\begin{tabular}{|c|c|}
\hline $\Delta \ln {\mathcal Z}$ & Evidence Result \\ 
\hline $0 - 1.0$ & Insignificant  \\ 
\hline $1.0 - 2.5$ &  Significant for higher {$\mathcal Z$} \\ 
\hline $2.5 - 5.0$ &  Strong for higher {$\mathcal Z$} \\ 
\hline  $ >  5.0$ &  Decisive for higher {$\mathcal Z$} \\ 
\hline 
\end{tabular}
\end{table}
\end{center}
\end{widetext} 

For computing {\it Bayesian Evidence}, we consider the \textsc{MultiNest}\cite{multinest} sampling algorithm using the python implementation \textit{pymultinest}\cite{pymultinest}.

\vspace{2mm}
In Table III, we show the $\ln({\mathcal Z})$ for different models considered, for four different combinations of cosmological data: BAO+TDSL+H(z), BAO+TDSL+H(z)+SNIa, BAO+TDSL+H(z)+SNIa+CMB and BAO+TDSL+H(z)+SNIa+CMB+GROWTH.

In Figure 1, we show the evidences for different models compared to the $\Lambda$CDM for four combinations of  cosmological data that we mention above. The horizontal line in each plot represents the $\Lambda$CDM model. Any model above this line has better evidence than $\Lambda$CDM whereas any model below this line has poorer evidence than $\Lambda$CDM. The number associated with each model represents its $\Delta \ln {\mathcal Z}$ from $\Lambda$CDM and this number determines how good/bad the model is, compared to $\Lambda$CDM, according to Jeffreys' scale in Table II.

For cosmological data from low-redshift measurements, e.g BAO, TDSL, H(z) and SNIa, all the thawing scalar field models (canonical scalar, tachyon and galileon) with different potentials has significantly better evidences than $\Lambda$CDM model. The tracking model with double exponential potential is indistinguishable from $\Lambda$CDM without the SNIa data. Inclusion of SNIa data makes the tracking model marginally better than $\Lambda$CDM. On the contrary, all the different dark energy parametrisations (except the GCG and the JBP) have significantly lower evidences than $\Lambda$CDM. The two GCG parametrisations ( GCG thawer and GCG tracker) are indistinguishable from $\Lambda$CDM. The JBP is significantly better than $\Lambda$CDM without the SNIa data whereas inclusion of SNIa data makes it indistinguishable from $\Lambda$CDM.

\begin{widetext}
\begin{center}
\begin{table}[H]
\centering
\caption{Log Evidence($\mathcal{Z}$)}
\begin{tabular}{lcccc}
\Xhline{2\arrayrulewidth}
Models & \multicolumn{1}{|c|}{\begin{tabular}[c]{@{}c@{}}BAO\\ +TDSL+H(z)\end{tabular}} & \multicolumn{1}{c|}{\begin{tabular}[c]{@{}c@{}}BAO+TDSL\\ +H(z)+\\ SN1a\end{tabular}} & \multicolumn{1}{c|}{\begin{tabular}[c]{@{}c@{}}BAO+TDSL\\ +H(z)\\ +SN1a+CMB\end{tabular}} & \multicolumn{1}{c|}{\begin{tabular}[c]{@{}c@{}}BAO+TDSL\\ +H(z)+SN1a\\ +CMB\\+GROWTH\end{tabular}}\\
\Xhline{2\arrayrulewidth}					
$\Lambda$ CDM &  -50.39 & -68.11 & -76.51 & -81.73\\
$wCDM$ & -51.59 & -69.27 & -78.45 & -84.13\\
$CPL$ & -52.06 & -69.73 & -79.09 & -84.7\\
$7CPL$ & -51.58 & -69.15 & -77.95 & -83.79\\
$BA$ & -51.44 & -69.8 & -78.99 & -84.54\\
$JBP$ & -48.72 & -68.39 & -79.15 & -84.86\\
Exponential Galileon & -48.7 & -66.37 & -75.54 & -80.71\\
Exponential Scalar & -48.64 & -66.38 & -74.11 & -79.57\\
Exponential Tachyon & -48.78 & -66.24 & -74.24 & -79.89\\
Inverse Galileon & -48.83 & -66.09 & -75.21 & -80.76\\
Inverse scalar & -48.7 & -66.04 & -74.31 & -79.69\\
Inverse Tachyon & -48.68 & -66.11 & -74.24 & -80\\
Linear Galileon & -48.65 & -66.38 & -75.33 & -80.97\\
Linear Scalar & -48.93 & -66.38 & -74.15 & -79.94\\
Linear Tachyon & -49.01 & -66.49 & -74.11 & -80.15\\
Square Galileon & -48.77 & -66.27 & -75.29 & -80.98\\
Square Scalar & -48.77 & -66.35 & -74.13 & -79.84\\
Square Tachyon & -48.94 & -66.17 & -74.31 & -79.55\\
GCG tracker &-50.85 & -68.46 & -80.32 & -85.85\\
GCG thawing &-50.86 & -68.43 & -78.09 & -83.49\\
Scalar Tracker & -49.49 & -67.03 &-76.28 & -81.63\\
\Xhline{2\arrayrulewidth} 
\end{tabular}
\end{table}
{\label{ln_e}}
\end{center}
\end{widetext}

Next we add the high redshift measurements from CMB with low redshifts data from BAO, TDSL, H(z) and SNIa. Addition of CMB data makes all the galileon models closer to $\Lambda$CDM compared to the previous case. The exponential galileon is now indistinguishable from $\Lambda$CDM whereas rest of the galileon models still have significantly better evidences compared to $\Lambda$CDM. On the other hand, all the canonical scalar and tachyon models have much better evidences from $\Lambda$CDM compared to the previous cases without CMB data.  The scalar tracker model is still indistinguishable from $\Lambda$CDM. GCG-tracker has now the lowest evidence among all models whereas GCG-thawer is significantly bad compared to $\Lambda$CDM. CPL parametrisation, which is the most studied parametrisation for dark energy models, is strongly disfavoured compared to $\Lambda$CDM and close to decisively disfavoured compared to canonical scalar and tachyon models. This is same for BA and JBP parametrisation. The wCDM and 7CPL models are still significantly disfavoured compared to $\Lambda$CDM. 

Finally we add the $f\sigma_{8}$ measurements from growth of large scale structures. All the galileon models are now indistinguishable from $\Lambda$CDM. The CPL, JBP, BA and GCG-tracker models now have even lower evidences compared to $\Lambda$CDM, canonical scalar and tachyon models. CPL and GCG-tracker have now decisively ruled out compared to canonical scalar and tachyon models.

To summarise, our analysis consistently shows that different scalar fields models which are non-phantom by definition have much better evidences than different dark energy parametrisations which allow both phantom and non-phantom behaviour, even if they have the same number of parameters. It is interesting to note that CMB and to some extent Growth data play a crucial role in disfavouring different parametrisations for dark energy equation of state compared to $\Lambda$CDM or different scalar field models.

\begin{figure*}[!htb]
\begin{center} 
\resizebox{200pt}{160pt}{\includegraphics{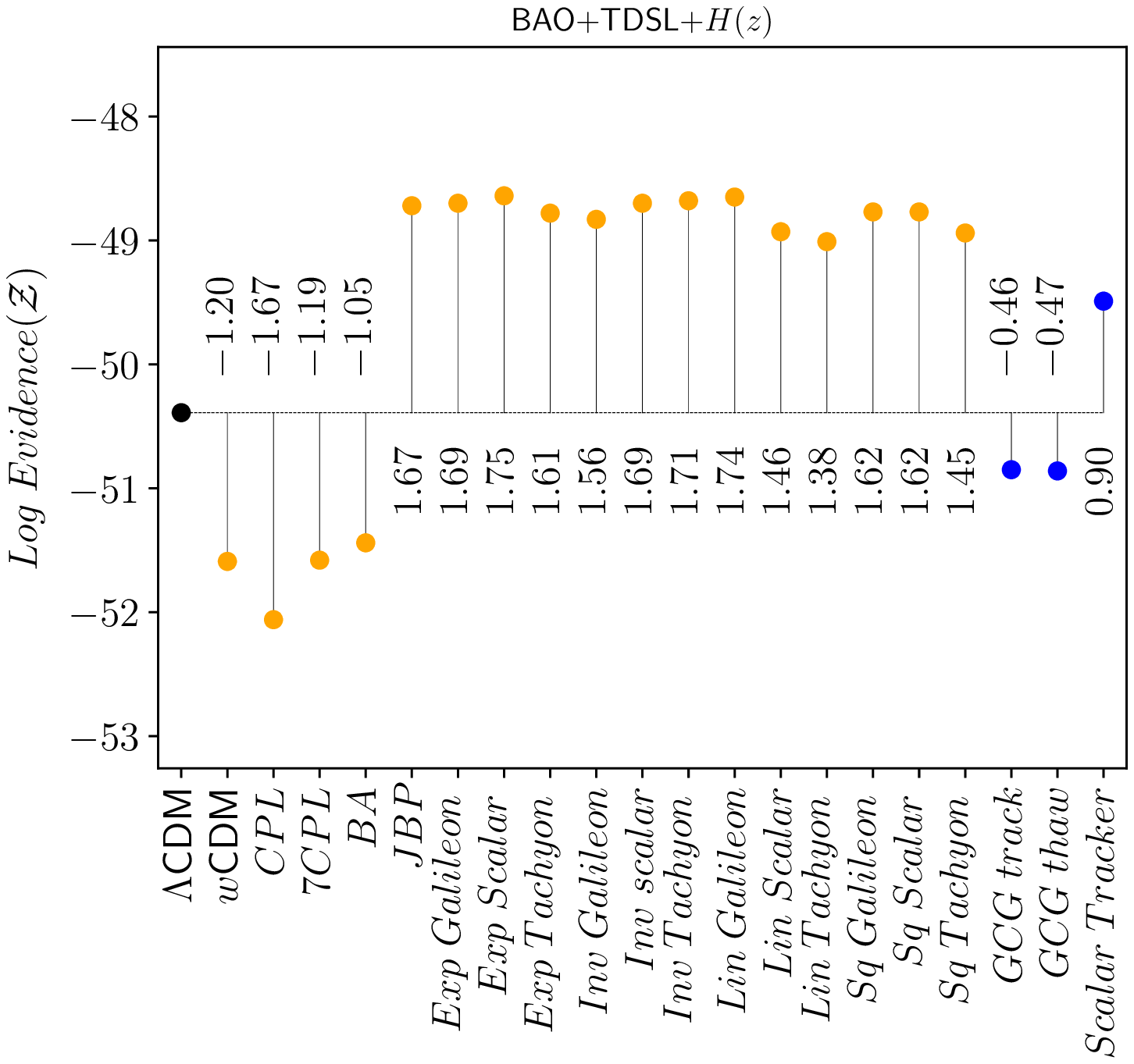}}
\resizebox{200pt}{160pt}{\includegraphics{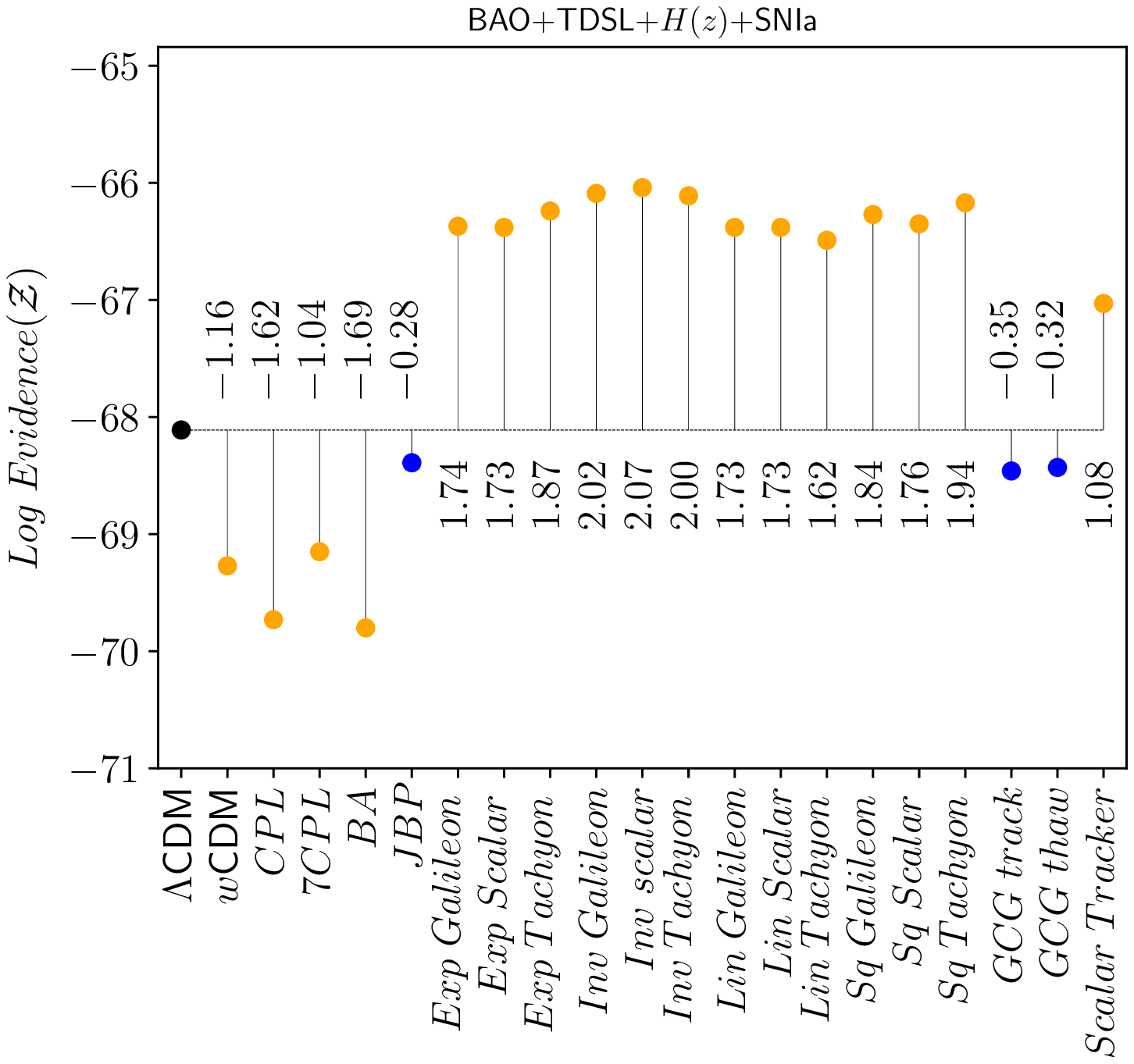}} \\
\hspace{1mm} \resizebox{200pt}{160pt}{\includegraphics{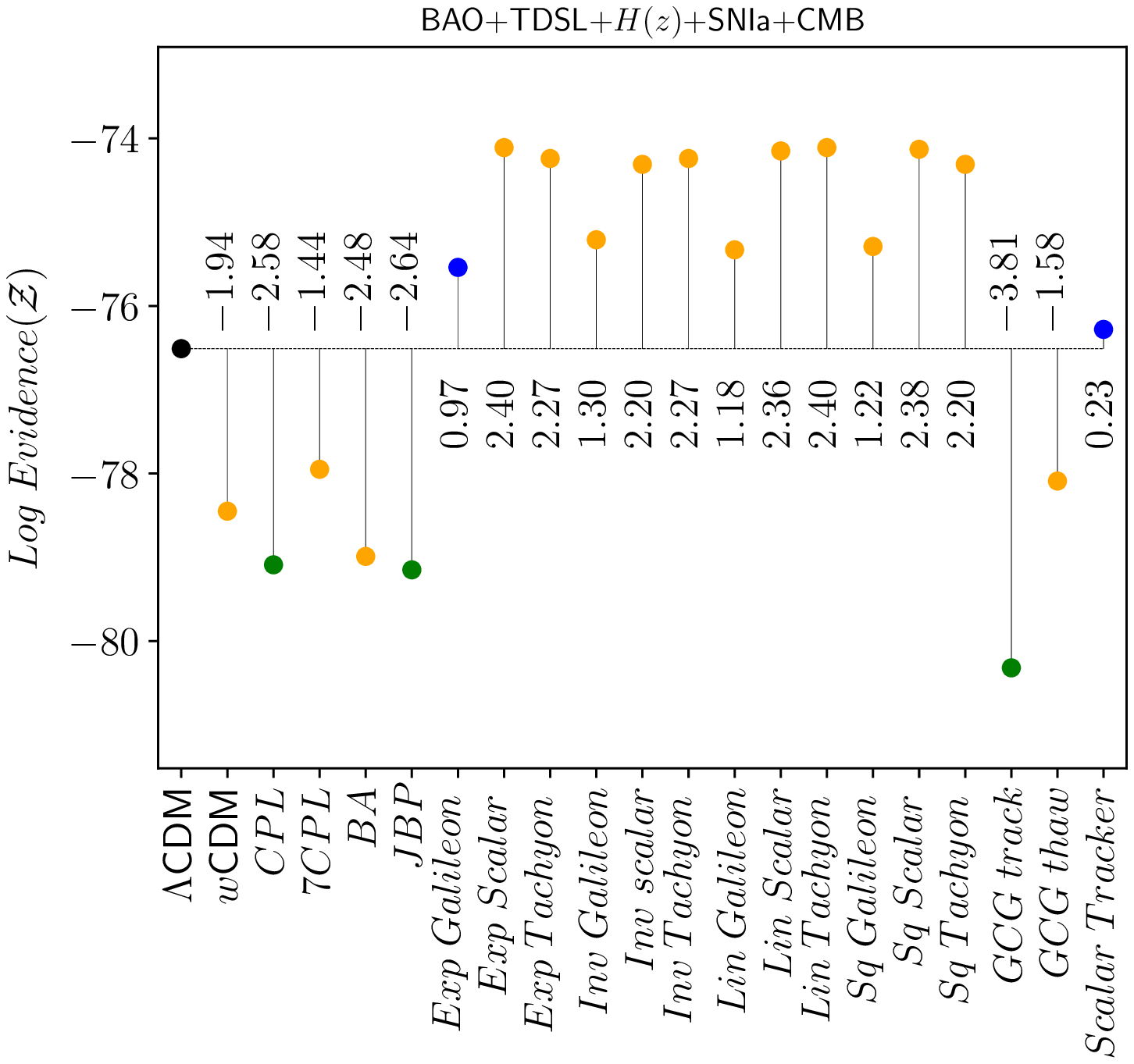}}
\hspace{1mm} \resizebox{200pt}{160pt}{\includegraphics{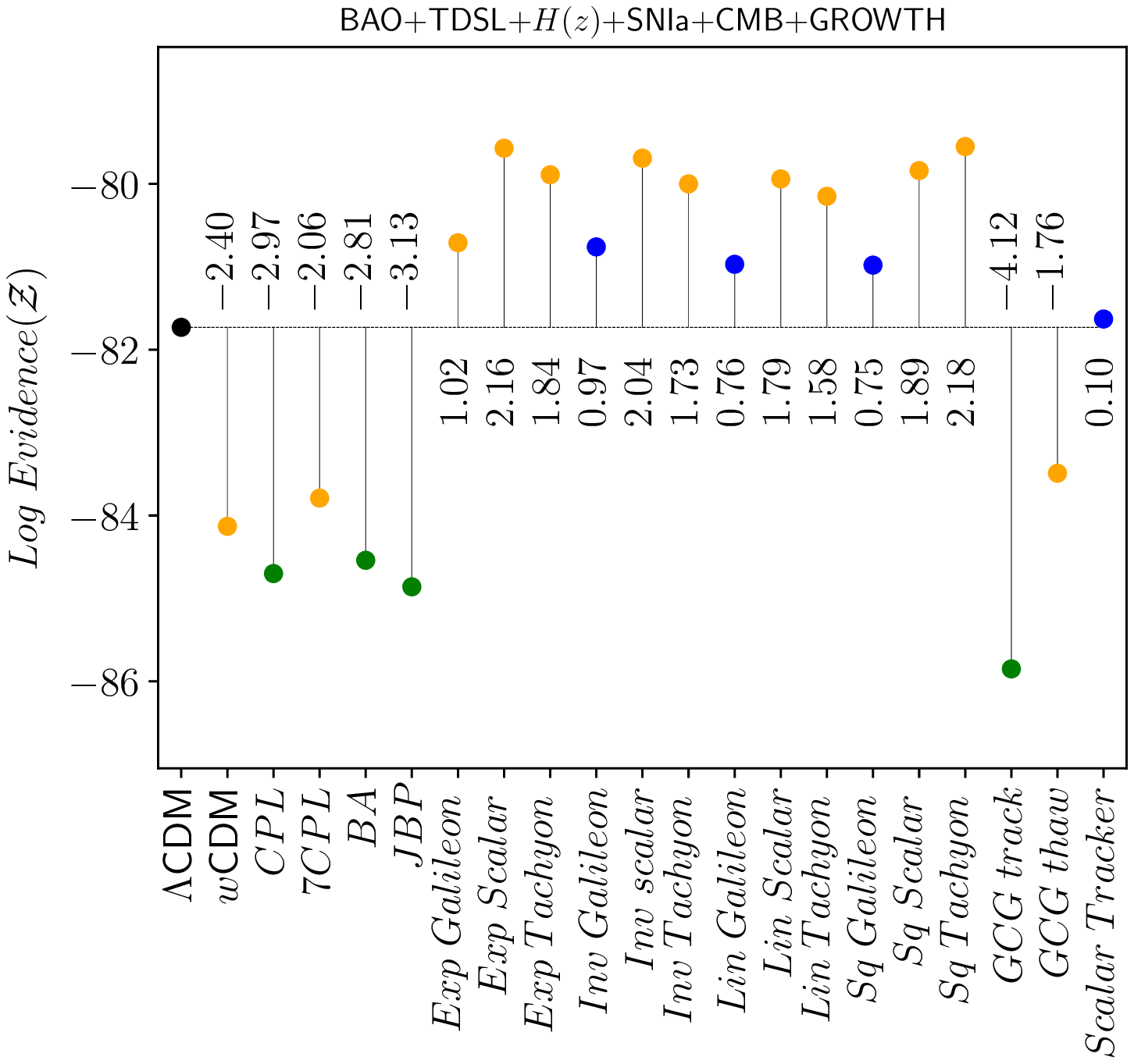}}
\end{center}
\caption{$\Delta \ln {\mathcal Z}$ for different models compared to $\Lambda$CDM for four combinations of observational data. The horizontal line represents the $\Lambda$CDM model. Models above this line have better evidences than $\Lambda$CDM whereas models below this have lower evidences than $\Lambda$CDM.  The blue, yellow and green points represent models with insignificant, significant and strong evidence compared to $\Lambda$CDM according to Jeffrey's scale given in Table II. } 
\end{figure*}
 
\section{Accelerating Vs. Non-Accelerating Universe} 
The first confirmation for a late time accelerating Universe was obtained through Type-Ia Supernova observations. But this has been questioned in recent times by Nielsen et al. \cite{nielsen} ( see also various discussions by Rubin and Hayden \citep{rubin}, Tutusaus {\it et al.}  \citep{tutusaus}, Dam {\it et al} \cite{dam}). In recent works, Ata {\it et al} \citep{ata} and Haridasu {\it et al} \citep{haridasu} have shown that with $\Lambda$CDM model, $\Omega_{\Lambda}=0$ non-accelerating model is ruled out at $6\sigma$ confidence level with BAO data alone, although BAO data alone can not provide a significant evidence for an accelerating universe in $w$CDM model \citep{hiradasu}.

In this section, we compare the {\it Bayesian evidences} for concordance $\Lambda$CDM model and a non-accelerated power law model as considered by Tutusaus {\it et al.} \citep{tutusaus}. The Hubble parameter for this non-accelerated model is given by:
\begin{equation}
H(z) = H_{0}(1+z)^{1/n},
\end{equation}

\begin{table}[!htb]
\centering
\caption{Log Evidence($\mathcal{Z}$) for $\Lambda$CDM Vs Non-Accelerating Universe}
\begin{tabular}{|c|c|c|c|c|c|c|}
\hline \rule[-2ex]{0pt}{5.5ex}  & BAO & +Masers+TDSL & +H(z) & +Growth & +$H_{0}$ & +SNIa\\ 
\hline \rule[-2ex]{0pt}{5.5ex} $\Lambda$CDM & -13.36 & -35.76 & -48.52 & -54.32 & -56.10 & -74.027\\ 
\hline \rule[-2ex]{0pt}{5.5ex} Non-Accelerating & -15.559 &  -38.57 &  -56.197 & -63.83 & -78.99 & -117.667\\ 
\hline 
\end{tabular} 
\end{table}

\noindent
where $n$ is a constant parameter. For a non-accelerated power law model for expanding universe, $0 < n \leq 1$. We consider all the low redshifts measurements as mentioned in section III. This is because the non-accelerating model given by equation (22) with a constant $n$ can not mimic at the same time a low redshift matter dominated Universe and a high redshift radiation dominated Universe. The results for evidence calculations for different combination of data are shown in Table IV. As one can see with BAO only data or with BAO+megamasers+TDSL data, there is no strong evidence for $\Lambda$CDM model as compared to non-accelerating power law model. But inclusion of $H(z)$, $H_{0}$, growth as well as SNIa measurements, results decisive evidence for $\Lambda$CDM model as compared to non-accelerating power law model.  Hence combination of different low-redshift data gives decisive evidence for late time accelerating universe.

\section{Conclusion}

We do a comprehensive study for {\it Bayesian Evidence} of dark energy models. We consider $21$ different dark energy models consisting both field theoretical models as well as different parametrisations for dark energy equation of state. And this is probably the highest number of dark energy models till date that has been considered for evidence calculations. We consider the data from BAO, TDSL, H(z) including the R16 measurement of $H_{0}$, SNIa, CMB, as well as the Growth measurements. We consider various combination of data and study how the evidence changes for these combinations.

The most significant result in our study is the fact that non-phantom models like scalar, galileon and tachyon models have always better evidences than dark energy parametrisations that allow both phantom and non-phantom behaviour. Although this evidence is significant to strong for BAO+TDSL+H(z)+SNIa data, inclusion of CMB and GROWTH data decisively rule out most of the dark energy parametrisations including the widely studied CPL one, compared to canonical scalar and tachyon models. As none of the seven dark energy parametrisations shows better evidence than different scalar field models, it raises the obvious question that whether using parametrised equation of state to model dark energy evolution to fit with observational data, is a good practice to study dark energy properties. In future, we need to focus more on actual field theoretical dark energy models rather than different parametrisations for dark energy equation of state or we need to have new parametrisations for dark energy equation of state that have better evidences compared to the existing ones. 

With the full set of data, all the galileon models as well as the scalar tracker model are indistinguishable from  $\Lambda$CDM. One needs to see whether future data can change this conclusion.

We also show that there is significant evidences in favour of scalar and tachyon models, compared to $\Lambda$CDM, but we do not find any model that is strongly or decisively favoured compared to $\Lambda$CDM. Given that we consider a very large class of dark energy models and consider a wide range of observational data, this shows that at present there is no strong or decisive evidence for dark energy evolution. This is consistent with recent findings about evidence for dark energy evolution \citep{alex}.

We also show that combination of different low redshift data itself can give decisive evidence for a accelerating universe although BAO measurements alone can not distinguish between $\Lambda$CDM and a non-accelerating power law model. This is in contrast to the recent claim by Ata {\it et al} \citep{ata} and Haridasu {\it et al} \citep{haridasu} that BAO measurements alone can rule out non-accelerating model at $5-6\sigma$ confidence level.

In near future, we shall extend our study on {\it Bayesian Evidence} to $f(R)$ and other modified gravity theories as well as to interacting dark energy models.

\section{Acknowledgement}
The author SK is funded by Scientific and Engineering Research Board, Department of Science and technology, Govt. of India through national post doctoral fellowship scheme. The authors Ruchika and BRD are funded by Council of Scientific and Industrial Research (CSIR), Govt. of India through Junior Research Fellowship and Senior Research Fellowship schemes Respectively. The author AL acknowledges the funding from DST SERB and research facilities provided at CTP, JMI during the work. The AAS acknowledges the financial support from IUCAA, Pune through its Associate Program where part of the work has been done. SK acknowledges computing facilities at ICTS-TIFR.

\end{document}